\documentstyle [12pt] {article}

\begin{document}

\baselineskip 7.5mm

\begin{flushright}
\begin{tabular}{l}
ITP-SB-92-38    \\
July 6, 1992
\end{tabular}
\end{flushright}

\vspace{8mm}
\begin{center}
{\Large \bf Invariants Describing Quark Mixing }\\
\vspace{4mm}
{\Large \bf and Mass Matrices} \\
\vspace{16mm}
  Alexander Kusenko\footnote{Supported in part by NSF contract PHY-89-08495} \\
\vspace{6mm}
Institute for Theoretical Physics  \\
State University of New York       \\
Stony Brook, N. Y. 11794-3840\footnote{email: kusenko@sunysbnp.bitnet}  \\

\vspace{20mm}

{\bf Abstract}
\end{center}

  We introduce two new sets of invariant functions of quark mass matrices,
which express the constraints on these mass matrices due to knowledge of the
quark mixing matrix.  These invariants provide a very simple method to test
candidate forms for mass matrices.

\vspace{50mm}

\pagestyle{empty}
\newpage

\pagestyle{plain}
\pagenumbering{arabic}

In recent years the increasingly accurate data on quark mixing has stimulated
one's interest in possible predictions concerning quark mass matrices.
The parameters of the Cabibbo-Kobayashi-Maskawa mixing matrix $V$
are now known
reasonably well \cite{data}.  This determination has been made possible partly
by the finding that there are only three generations of usual standard-model
fermions (with corresponding light or massless neutrinos).  Since the
diagonalization of the quark matrices in the up and down sectors determines
$V$, one can work back from the knowledge of $V$ to put constraints on the
possible forms of (original, nondiagonal) quark mass matrices.  However,
the data on quark mixing determines these mass matrices
only up to an arbitrary unitary similarity transformation.  This is a result of
the fact that if the up and down
quark mass matrices, $M_{u}$ and $M_{d}$, are both acted on by the same
unitary operator $U_{0}$ according to

\begin{equation}
M_{u,d} \ \rightarrow \ U_{0} \: M_{u,d} \: U_{0}^{\dag}  \label{u0}
\end{equation}
then the mixing matrix $V$ remains unchanged.
There have been many attempts to study specific assumed forms for quark
matrices.  While this is worthwhile, it is desirable to
express the constraints from data on $V$  on the quark mass matrices in an
invariant form.  In Refs. \cite{Jarlskog,Branco}  certain invariant
functions of the quark mass matrices
$I_{pq}$ were introduced, which are expressed in terms of the quark masses
squared and the $|V_{ij}|$:
\begin{equation}
I_{pq} = Tr(H_{u}^{p} \: H_{d}^{q}) =\sum_{ij} \: (m^{(u)}_{i})^{2 p}
(m^{(d)}_{j})^{2 q} \: |V_{ij}|^{2}
\label{ieq}
\end{equation}
where $H_{q}=M_{q} M_{q}^{\dag}$, and $m_{i}^{(u)}$ and $m_{i}^{(u)}$ are
the masses of quarks in the ``up'' and ``down'' charge sectors.

However, in practice, these are somewhat awkward
to use because of the high powers of elements of the actual quark mass matrices
which are involved.

In this paper we introduce two new sets of invariants of the quark mass
matrices (with respect to the transformation (\ref{u0})) which can be
 expressed in terms of the measurable quantities only  (up to the $\pm$ sign
ambiguity in fermion masses, which, as we will show, can be dealt with by
considering all choices of these signs).  These invariants have an important
advantage relative to (\ref{ieq}) that they involve lower powers of the
elements of the quark mass matrices and hence yield much less complicated
analytic expressions in practice.

In the standard model with only left-handed charged weak currents one may
choose $M_{u}$ and $M_{d}$ both to be hermitian without any loss of
generality, by re-phasing the right-handed components of quarks.

We introduce the following new invariants of the transformation (\ref{u0}):

\begin{equation}
K_{pq}=Tr ( \: M_{u}^{p} \:  M^{q}_{d} \:  )   \label{I}
\end{equation}

\begin{equation}
L_{pq}(\alpha,\beta)= det ( \alpha M_{u}^{p}+ \beta M_{d}^{q} ) \label{J}
\end{equation}
where $ p,q,\alpha, \beta  \ \neq 0 $.

 The hermitian matrices $M_{u}$
and $M_{d}$ can be diagonalized by a unitary similarity transformation:

\begin{eqnarray}
\left \{ \begin{array}{l}
         U_{u} M_{u} U_{u}^{\dag} = D_{u} \\  \\
         U_{d} M_{d} U_{d}^{\dag} = D_{d}
         \end{array} \right.
\end{eqnarray}
where $D_{q}=diag(m_{1}^{(q)},m_{2}^{(q)},m_{3}^{(q)})$ are the
diagonal matrices of the quark masses.

The mixing matrix $V$ can be written then as:

\begin{equation}
V= U_{u} U_{d}^{\dag}
\end{equation}

We can now express the invariants $K_{pq}$ in terms of the squares of
absolute values of the elements of the mixing matrix $ U_{ij}=|V_{ij}|^2$
(the latter are measurable quantities \cite{data}):

\begin{equation}
\begin{array}{c}
K_{pq}=Tr (U_{u}^{\dag} D_{u}^{p} U_{p} U_{d}^{\dag} D_{d}^{q} U_{d})=
Tr(V^{\dag} D_{u}^{p} V D_{d}^{q})=  \\  \\
\sum_{ij} \ (m_{i}^{(u)})^{p} \ (m_{j}^{(d)})^{q} \: U_{ij}
\end{array}  \label{I_U}
\end{equation}

In order to find an expression for $L_{pq}$ in terms of $U_{ij}$ we will need
the following

\pagebreak

{\em Theorem  1.}

{\em If $A$ and $B$ are two $ 3\times 3$ matrices such that $ det(A) \neq 0$
and $ det(B) \neq 0$ then the following relation holds:}

\begin{equation}
det(A+B)= det(A) + det(B) + det(A) \: Tr(A^{-1} B)+ det(B) \: Tr(A B^{-1})
\label{theorem}
\end{equation}

\vspace{4mm}

{\em Proof:}

We denote the elements of matrices $A$ and $B$ by $A_{ij}$ and $B_{ij}$
correspondingly.  Their co-factors (which are equal to the corresponding
minors, up to sign) will be written as $\hat{A}_{ij}$ and $\hat{B}_{ij}$.
Then each determinant may be decomposed in a sum (Laplace expansion):

\[
det(A)= \sum_{i} A_{ij} \hat{A}_{ij}=\sum_{j} A_{ij} \hat{A}_{ij}  \]
\[ det(B)= \sum_{i} B_{ij} \hat{B}_{ij}=\sum_{j} B_{ij} \hat{B}_{ij}
\]

By definition, the determinant of a $3\times 3$ matrix is a sum of $n!$ terms:

\begin{equation}
det(A+B)= \sum (-1)^{r} (A_{1 k_{1}} + B_{1 k_{1}} )
(A_{2 k_{2}}+ B_{2 k_{2}} ) (A_{3 k_{3}} + B_{3 k_{3}})  \label{det}
\end{equation}
where $r$ is the sign of the permutation $(^{1}_{k_{1}} \
^{2}_{k_{2}} \ ^{3}_{k_{3}})$.

The terms in the sum (\ref{det}) which contain only the elements of $A$
can be arranged as $det(A)$.  Similarly, the terms containing only
$B$'s give $det(B)$.  The terms containing one element of $A$ multiplied by
two elements of $B$, or visa versa, can be rewritten as:

\begin{equation}
\sum_{i,j} (A_{ij} \hat{B}_{ij}+B_{ij} \hat{A}_{ij})  \label{ab}
\end{equation}

We can now use an identity:

\[ (A^{-1})_{ij}= \frac{1}{det(A)} \hat{A}_{ji}  \]
to rewrite (\ref{ab}) as:

\[
 \sum_{i,j} (A_{ij} \hat{B}_{ij}+B_{ij} \hat{A}_{ij})
= det(B) \: \sum_{ij} A_{ij} (B^{-1})_{ji}+
det(A) \: \sum_{ij} (A^{-1})_{ji} B_{ij}=   \]

\[ det(A) \: Tr(A^{-1} B)+
det(B) \: Tr(A B^{-1})  \]

All together we get

\[ det(A+B)= det(A) + det(B) + det(A) \: Tr(A^{-1} B)+ det(B) \:
Tr(A B^{-1}) \]
which is the statement of {\em Theorem 1}.  This completes the proof.

  {\em Theorem 1} may be easily generalized to the case of
$2\times 2$ matrices, in which case the last two terms in (\ref{theorem})
are equal and correspond
to a redundant counting of the same terms in a sum similar to (\ref{det}).
Thus for the $2\times 2$ matrices we get:

\begin{equation}
\begin{array}{c}
det(A+B)= det(A) + det(B) + det(A) \: Tr(A^{-1} B) \equiv \\  \\
 det(A) + det(B) + det(B) \: Tr(A B^{-1})
\end{array}
\end{equation}

The immediate consequence of {\em Theorem 1} and equation (\ref{I_U})
is the following relation:

\begin{eqnarray}
L_{pq}(\alpha, \beta) \ \equiv \ det(\alpha M_{u}^{p}+\beta M_{d}^{q}) \ =
\ \ \ \ \ \ \ \ \ \ \ \ \ \ \ \ \ \ \ \ \ \ \ \ \ \   \nonumber \\ \nonumber \\
\alpha^{3} \ (m_{1}^{(u)} m_{2}^{(u)} m_{3}^{(u)})^{p} \
[\: 1+ (\beta/\alpha) \sum_{ij} \: [(m_{j}^{(d)})^{ q}/(m_{i}^{(u)})^{ p}]\:
U_{ij} \: ]   +  \label{idm} \\ \nonumber \\
\beta^{3} \ (m_{1}^{(d)} m_{2}^{(d)} m_{3}^{(d)})^{q} \
[\: 1+  (\alpha/\beta) \sum_{ij} \: [(m_{j}^{(u)})^{ p}/(m_{i}^{(d)})^{ q}] \:
U_{ij} \: ]   \nonumber
\end{eqnarray}

We also notice that

\begin{equation}
L_{pq}(1, \pm 1) =
(m_{1}^{(u)} m_{2}^{(u)} m_{3}^{(u)})^{p} (1 \pm K_{(-p) \, q}) \pm
(m_{1}^{(d)} m_{2}^{(d)} m_{3}^{(d)})^{q} (1 \pm K_{p \, (-q)}) \label{ij}
\end{equation}

In summary, we have introduced and studied two new sets of invariants of
the quark mass matrices which are model- and weak basis-independent.  The
identities (\ref{idm}) and (\ref{ij})
involving these invariants which we have derived above
provide important constraints on the possible
 forms of quark mass matrices $M_{u}$ and $M_d$
since they directly relate the elements of these matrices to
the measurable parameters $|V_{ij}|^2$ and quark masses, and thereby enable one
to avoid the explicit calculation of the eigenvectors of $M_{u}$ and $M_{d}$.
(For another approach concerned with obtaining information on these mass
matrices from $V$, see, e.g., Ref. \cite{ak}.)
Although these invariants involve ambiguities stemming from the fact that the
signs of fermion masses are not physical, these ambiguities can be
dealt with by considering all possible sign combinations.  In practice, we have
found that this is not a serious complication.
Applications of these invariants will be discussed as part of a separate
paper \cite{next}.

The author is grateful to Professor Robert E. Shrock for
helpful discussions and for his comments.

\end{document}